# Quantum valley Hall states in low-buckled counterparts of graphene bilayer

Yu-Hao Shen[1,2], Jun-Ding Zheng[3,5], Wen-Yi Tong[3,5], Zhi-Qiang Bao[3,5], Xian-Gang Wan[1,2]*, Chun-Gang Duan[3,4,5]*

[1]National Laboratory of Solid-State Microstructures and School of Physics, Nanjing University, Nanjing 210093, China

[2]Collaborative Innovation Center of Advanced Microstructures, Nanjing University, Nanjing 210093, China

[3]Key Laboratory of Polar Materials and Devices, Ministry of Education, East China Normal University, Shanghai 200241, China

[4]Collaborative Innovation Center of Extreme Optics, Shanxi University, Taiyuan, Shanxi 030006, China.

[5]Shanghai Center of Brain-inspired Intelligent Materials and Devices, East China Normal University, Shanghai 200241, China

*Author to whom any correspondence should be addressed.
E-mail:
cgduan@clpm.ecnu.edu.cn
xgwan@nju.edu.cn

# ABSTRACT

With low-buckled structure for each layer in graphene bilayer system, there breaks inversion symmetry ($\mathcal{P}$-symmetry) for one stacking when both A and B sublattices in top layer are aligned with those in bottom layer. In consideration of spin-orbit coupling (SOC), there opens nontrivial topological gap in each monolayer system to achieve quantum spin Hall effect (QSHE). As long as time-reversal symmetry ($\mathcal{T}$-symmetry) is preserved, the gapless edge states is robust in each individual layer even for the bilayer absent of $\mathcal{PT}$ symmetry. Based on this platform and through tight-binding (TB) model calculations, we find that it evolves into a prototypical system capable of demonstrating the quantum valley Hall effect (QVHE) with introduction of a layer-resolved Rashba SOC, which induces band inversion at each K valley within the hexagonal Brillouin zone (BZ). The topological transition comes from that the valley Chern number $C_v = C_K - C_{K'}$ switches from 0 to 2, which characterizes the nontrivial QVHE phase transited from two coupled $Z_2$ topological insulators. We also demonstrate that the layer-resolved Rashba SOC can be introduced equivalently by twisting two van der Waals touched layers. And through TB calculations, it is shown that the K bands inverts in its corresponding mini BZ when the two layers twisted by a small angle. Our findings advance potential applications for the devices design in topological valleytronics and twistronics.

# INTRODUCTION

Spin-orbit gap opening in graphene system make it become ideal platforms for the achievement of nontrivial topological edge states[1-3]. It hosts QSHE phase with the time-reversal symmetry protected electron states gapped within the bulk and gapless states without dissipation that conducts spin at the sample boundaries. The low-buckled counterparts of graphene system e.g. silicene, germanene and stanene, are demonstrated to possess large spin-orbit gap[4,5]. Compared with the planar graphene, where QSHE can only occur at unrealistic low temperature, silicene family can sustain experimentally observable edge states, which is both topological nontrivial and robust metallic[6].

Moreover, valley degree of freedom, recognized as a kind of pseudospin in spin-orbit coupled honeycomb system[7-9], of the edge electron states can also be filtered like spin filtering in QSHE[10-13]. Its participation lies in that there opens a band gap at the Dirac points K and K' in such system and topological transition can occur for valley states i.e., the gap closes and reopens, respond to extrinsic Rashba SOC tuned by external field[12-14]. Thus in each valley subspace (K and K' valley states are decoupled as intervalley scattering ignored), there gives quantum anomalous Hall states or not[15,16], characterized by the nonzero or zero Chern number $C_{K/K'}$ respectively and could lead to the nonzero valley Chern number $C_v = C_K - C_{K'}$ which further characterizes QVHE. In a typical graphene bilayer[17], it is demonstrated that there exhibits topological transition from $C_v = 1$ phase to $C_v = 2$ phase whereas the topological invariant $Z_2$ changes from 0 to 1, meaning that both spin and valley filtering of the conductive edge modes exists in the latter case. It is noteworthy in that work[17] the external electric field is along out-of-plane direction on the AB-stacked bilayer. This indicates the $\mathcal{P}$-symmetry is absent, which motive us to design an AA-stacked graphene bilayer system with low-buckling in each layer and investigate both intrinsic and extrinsic SOC induced band inversion at each K valley.

In this paper, based on TB model calculations towards the AA-stacked bilayer system with low-buckled atomic layer. We show that with a layer-resolved Rashba SOC considered there give rise to band inversion at each K valley as the interlayer hopping parameters varies. There exhibits QVHE phase transition from $C_v = 0$ to $C_v = 2$ phase but $Z_2$ topological invariant is always vanished since even number of individual QSHE layers renders that back-scattering between counter-propagating channels at the same edge is not forbidden and spin accumulation can be cancelled. Furthermore, we point that the layer-resolved SOC effect, as an extrinsic Rashba SOC, can be equivalently introduced by intrinsic one in each buckled layer when the two van der Waals separated layers twisted and forms moire pattern. To demonstrate that, we perform TB approximation calculations based on the parameters fitted from first-principles calculated band structure. Here we choose germanene bilayer as an example. Our calculated results show the K bands can be inverted in its corresponding mini BZ as we twist. And we also calculate its topological nontrivial edge states for a nanoribbon structure.

## TIGHT-BINDING MODEL

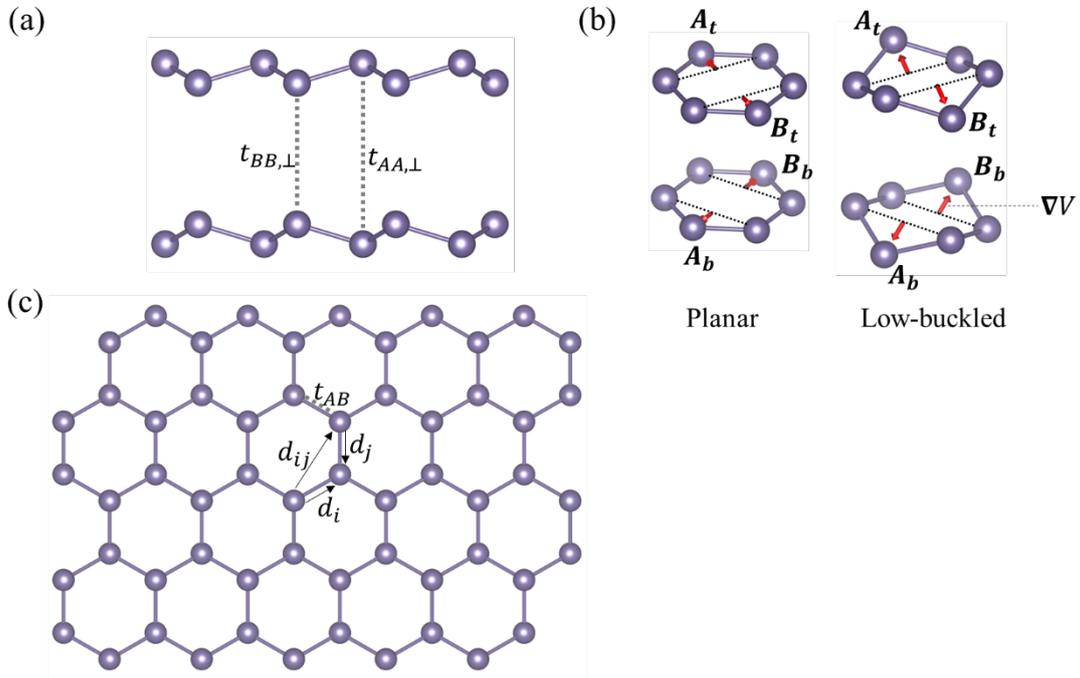

**Figure**. 1 (a) Side view of the AA stacked bilayer structure with low-buckling in each layer. (b) Comparison of the bilayer between planar and low-buckled layer structure. The dashed line connects intra-sublattice sites and red arrows denotes the direction of an effective electric field $\nabla V$ due to the non-centrosymmetric structure in each layer. (c) Top view of the honeycomb lattice in our bilayer system.

We start from a model Hamiltonian based on TB approximation of the bilayer system for the silicene family. As shown in Fig. 1, we choose AA stacking for the atomic structure without inversion symmetry, which possess D3h point group symmetry with respect to its hollow center of the honeycomb lattice. The intralayer Hamiltonian in the presence of SOC can be written as:

$$H_0 = -t \sum_{\langle ij \rangle, \alpha} c^\dagger_{i,\alpha} c_{j,\alpha} + i\lambda_{SOC} \sum_{\langle\langle ij \rangle\rangle, \alpha\beta} v_{ij} c^\dagger_{i,\alpha} s^z_{\alpha\beta} c_{j,\beta}$$

$$-i\lambda'_{SOC} \sum_{\langle\langle ij \rangle\rangle, \alpha\beta} l_{ij} \mu_{ij} c^\dagger_{i,\alpha} \left( s \times \hat{d}_{ij} \right)^z_{\alpha\beta} c_{j,\beta} + i\lambda''_{SOC} \sum_{\langle ij \rangle, \alpha\beta} l_{ij} c^\dagger_{i,\alpha} \left( s \times \hat{d}_{ij} \right)^z_{\alpha\beta} c_{j,\beta} \quad (1)$$

Here, $c^\dagger_{i,\alpha}$ ($c_{i,\alpha}$) is the usual creation operator for electron with spin $\alpha$ on site $i$ and $l_{ij} = \pm 1$ for top ($t$) and bottom ($b$) layer. In the first term, the hopping energy $t_{AB}$ between nearest neighbor sites denoted as $\langle ij \rangle$ is $t$. The intrinsic SOC terms with strength $\lambda_{SOC}$ and $\lambda'_{SOC}$ corresponds to the effective electric field $\nabla V$ origin from symmetry consideration, as shown in Fig. 1(b). They involve coupling energy between next-nearest neighbor sites denoted as $\langle\langle ij \rangle\rangle$ and $s$ are spin Pauli matrices. $v_{ij} = \dfrac{d_i \times d_j \cdot \hat{z}}{|d_i \times d_j \cdot \hat{z}|}$ with lattice vector $d_i$, $d_j$ and $d_{ij} = d_i - d_j$, as shown in Fig. 1(c) and $\mu_{ij} = \pm 1$ for A and B sublattice. The last term, which is extrinsic SOC for each layer, represents the layer-resolved Rashba coupling with strength $\lambda''_{SOC}$, where the lattice vector $d_{ij}$ connects nearest neighbor sites. The interlayer Hamiltonian is considered as:

$$H_1 = \sum_{\langle ii' \rangle, \langle jj' \rangle, \alpha} t_{AA,\perp} c^\dagger_{i,\alpha} c_{i',\alpha} + t_{BB,\perp} c^\dagger_{j,\alpha} c_{j',\alpha} \quad (2)$$

where $t_{AA,\perp} = m_0 + m_1$ and $t_{BB,\perp} = m_0 - m_1$. Here, $\langle ii' \rangle$ involve only the vertical hopping terms and the planar ones are ignored at present. For the case of interlayer distance much larger than the intralayer buckling, we also ignore the interlayer SOC contribution.

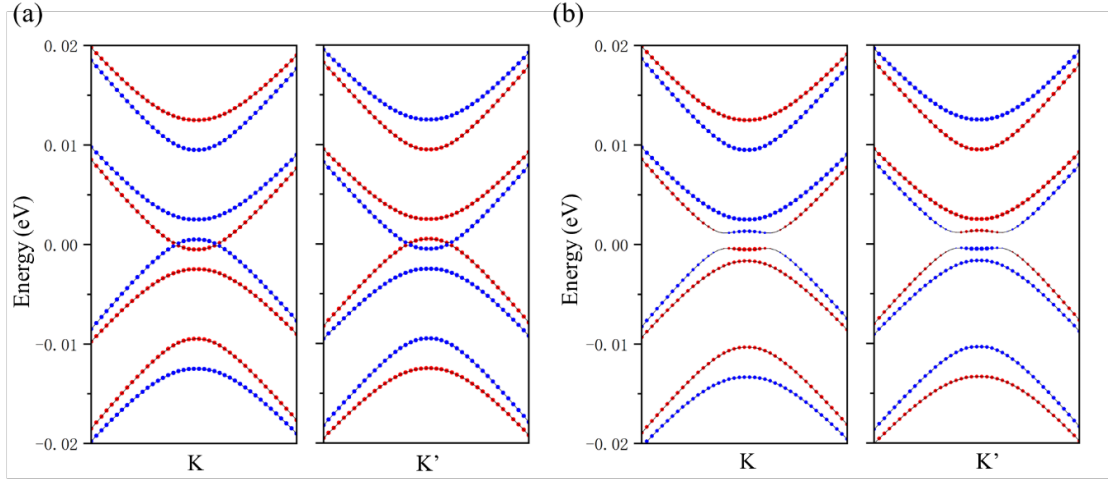

**Figure. 2** The calculated band structures in the vicinity of K and K' with (b) and without (a) layer-resolved Rashba SOC turned on. The red (blue) color shows the calculated positive (negative) spin $\langle s_z \rangle$ polarization on each $k$. The Fermi level is set up to 0 and the model parameters are chosen to be $t$=173.2, $\lambda_{SOC}$=1.2, $\lambda'_{SOC}$=2, $\lambda''_{SOC}$=1, $m_0$=5 and $m_1$=1.5, in units of meV.

## RESULTS AND DISCUSSION

### A. Topological phase transition in bilayer

Now the band structures can be obtained from above TB Hamiltonian $H = H_0 + H_1$ on the basis of $\{|A_t\rangle, |B_t\rangle, |A_b\rangle, |B_b\rangle\}^T \otimes \{|\uparrow\rangle, |\downarrow\rangle\}^T$. In Fig. 2 we compare the calculated band structures in the vicinity of K with $\lambda''_{SOC}$ (panel (b)) and without $\lambda''_{SOC}$ (panel (a)) turned on. For the case of the model parameters we choose at present, it is shown that the layer-resolved Rashba SOC results in band inversion at K and K' and the bilayer system become insulating, yet without it turned on the system is metallic. We even introduce planar interlayer coupling terms i.e., the hopping between nearest $A_t (B_t)$ and $B_b (A_b)$ sites into the Hamiltonian H and find it cannot open the bulk band gap when $\lambda''_{SOC}$ is turned off and there just breaks the particle-hole symmetry of the band structure.

The low energy physics can be captured by **k·p** effective model expanded around K and K' valley with deviated k to linear order. Because of horizontal mirror symmetry we diagonalize the Hamiltonian in bonding and antibonding of layer orbital basis $\{|A_t+A_b\rangle, |A_t-A_b\rangle, |B_t+B_b\rangle, |B_t-B_b\rangle\}^T \otimes \{|\uparrow\rangle, |\downarrow\rangle\}^T$, defined as sub-orbital space, and there gives:

$$H_\tau = \hbar v_F \mu_0 (\tau k_x \sigma_x + k_y \sigma_y) s_0 + m_0 \mu_z \sigma_0 s_0 + m_1 \mu_z \sigma_z s_0$$
$$+ \tilde{\lambda}_{SOC} \mu_0 \tau \sigma_z s_z + \tilde{\lambda}'_{SOC} \mu_x \tau \sigma_z (k_x s_y - k_y s_x) + \tilde{\lambda}''_{SOC} \mu_x (\tau \sigma_x s_y - \sigma_y s_x) \quad (3)$$

where $\tau = \pm 1$ is the valley index. **μ**, **σ** and **s** represents Pauli matrices refers to sub-orbital, sublattice and spin index, respectively. The model parameters become that $v_F = \sqrt{3} t / 2$, $\tilde{\lambda}_{SOC} = 3\sqrt{3} \lambda_{SOC}$, $\tilde{\lambda}'_{SOC} = 3 \lambda'_{SOC} / 2$ and $\tilde{\lambda}''_{SOC} = 3 \lambda''_{SOC} / 2$. Note that the valley K and K' remains good quantum number since intervalley scattering is absent. For inversion operation $\mathcal{P}$ in such basis of each valley $\tau$ space, we find $\mathcal{P} = \mu_z \sigma_x$ due to the exchange of both sublattices and layers. And the time-reversal operation $\mathcal{T} = \mathcal{K} s_y$. Thus, at each K (**k** = 0), the third term leads to $\mathcal{PT}$-symmetry breaking which lift the spin degeneracy. The last one, in terms of an extrinsic Rashba SOC, accounts for the band inversion, where it indeed couples the two sublattices at K. It is necessary for the band inversion occurs at K because of opposite polarized sublattice states for valence and conduction K band edges if extrinsic Rashba SOC $\lambda''_{SOC}$ turned off.

Moreover, the occurrence of the K band inversion suggests topological phase transition to QVHE insulator as long as the Fermi level lies inside the bulk energy gap. It can be achieved by the evolution of $m_0$ and $m_1$ at fixed other model parameters. Analytically, we find that the transition boundary for that is

$$\left( \hat{\lambda}_{SOC} \pm m_1 \right)^2 = 4 \hat{\lambda}''^2_{SOC} + m_0^2 \quad (4)$$

where ± indicates that there will occur twice phase transition as $m_0$ evolves at fixed both $m_1$ and other model parameters. However, nontrivial topological phase for each valley occurs only if bulk gap closing and reopening once, which should be reflected by a nonzero Chern number for each valley space. As we know, in absent of $\mathcal{T}$-symmetry, there possess quantum anomalous Hall states of a two-dimensional (2D) Chern insulator when the Fermi level lies inside the gap, characterized by a nonzero Chern [18]number $C$ as[7,18]:

$$C = \frac{1}{2\pi} \sum_{n \in occ} \oint d\mathbf{k} \cdot A_n(\mathbf{k}) = \frac{1}{2\pi} \sum_{n \in occ} \iint d^2 k \Omega_{n,z}(\mathbf{k}) \tag{5}$$

where Berry connection in $\mathbf{k}$ space is $A_n(\mathbf{k}) = i\langle u_n(\mathbf{k})|\nabla_\mathbf{k} u_n(\mathbf{k})\rangle$ for the $n$th band eigen states $|u_n(\mathbf{k})\rangle$ and the integral of the Berry curvature $\Omega_{n,z}(\mathbf{k}) = \nabla_\mathbf{k} \times A_n(\mathbf{k})$ is over the entire 2D BZ. The Chern number is calculated from the summation of this integral over all occupied bands. It is found that as the band inversion occur once, for K and K' valley states related by time reversal symmetry, the Chern number contribution is $C_K = 1$ and $C_{K'} = -1$ by integrating over the neighborhood about K and K', respectively. And $C_K = C_{K'} = 0$ in the case of no band inversion and twice occurrence. Then we have a nonzero valley Chern number $C_v = C_K - C_{K'} = 2$, characterizing the QVHE phase.

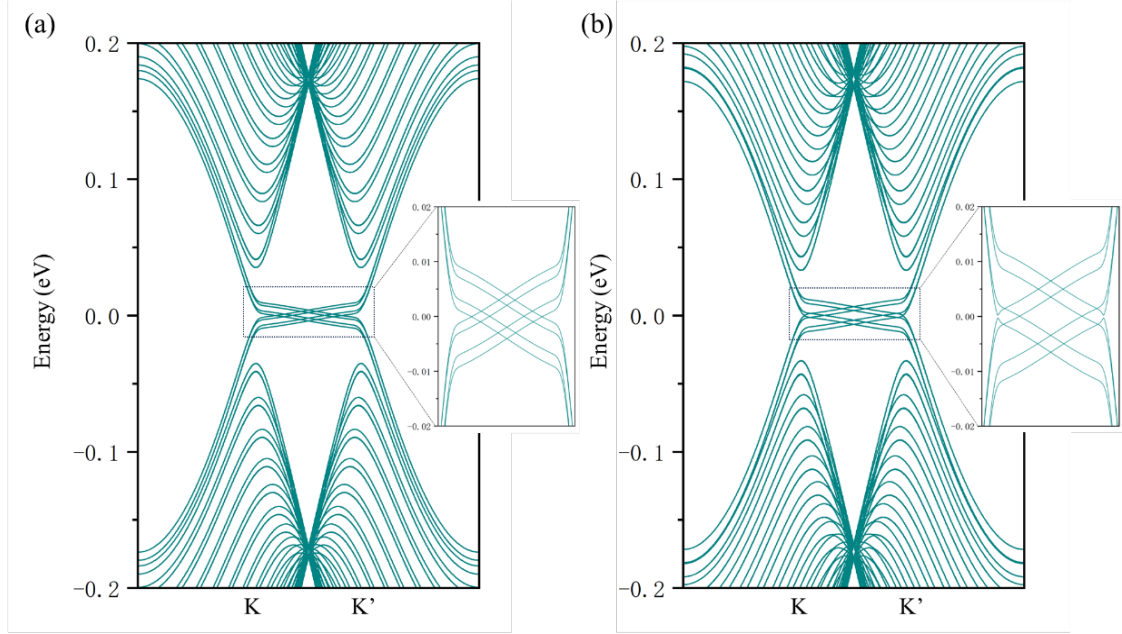

**Figure. 3** Calculated band structure of one-dimensional periodic nanoribbon structure with zigzag termination. The model parameters are chosen to be (a) $m_0=3$, $m_1=1$ (b) $m_0=5$, $m_1=1.5$ in units of meV. Other fixed parameters are same as those in the bulk band structures calculations.

The topological phase transition can be reflected by bulk gap closing and reopening and the corresponding helical edge states which cross the Fermi level (set to be zero) will be gapped when the system evolves from two coupled QSHE layers. As shown in the edge band structure we calculate in Fig. 3, there left one Kramer's pair of gapless edge states as transition to QVHE phase (Fig. 3(b)). Note that for a nanoribbon structure with zigzag or armchair terminations, the K and K' valley states can remain decoupled in zigzag-edged strips but are coupled in armchair-edged strips. For the latter case, there is no QVHE phase and through calculating its edge states in the corresponding nanostructure, we find no gapless states exist before and after the bulk gap closing and reopening. It is indicated that the bilayer system could not be a strong topological insulator but only can be distinguished from a trivial insulator by the existence of the quantized valley Hall conductance, characterized by the valley Chern number $C_v = 2$. As for its helical edge states there is no spin-filtering. It can be further revealed by examining the $Z_2$ topological invariant calculated as[19]:

$$Z_2 = \left( \frac{1}{2\pi} \sum_{n \in occ} \oint d\mathbf{k} \cdot \mathbf{A}_n(\mathbf{k}) - \iint d^2\mathbf{k}\, \Omega_{n,z}(\mathbf{k}) \right) \bmod 2 \tag{6}$$

which characterize the topological obstruction in the presence of $\mathcal{T}$-symmetry constraint for the eigen states $|u_n(-\mathbf{k})\rangle = \mathcal{T}|u_n(\mathbf{k})\rangle$. Here, the Berry connection $\mathbf{A}_n(\mathbf{k})$ and Berry curvature $\Omega_{n,z}(\mathbf{k})$ integral is along a closed boundary of half BZ and over the half BZ respectively. We find that $Z_2$ is always vanished for the two coupled QSHE layers in our case. For the case before bulk gap closing (Fig. 3(a)), even number of Kramer's pairs of the gapless edge states cannot be robust against the back-scattering from disorder that even exist in the presence of $\mathcal{T}$-symmetry. And in the case after bulk gap reopening (Fig. 3(b)), although one pair is gapped, the spin accumulation is always vanished. As a consequence, along the zigzag sample boundary, there only filters valley states via a pure valley current i.e., two counter propagating edge states that are time reversal symmetry related, similar to the case of Kramer's pairs protected by $\mathcal{T}$-symmetry in QSHE phase.

## B. Topological transition induced by interlayer twist

The band topology of silicene family is robust against the intrinsic Rashba SOC $\lambda'_{SOC}$ due to the low-buckling of the atomic layer, which vanishes at K and K'. This is entirely different from the extrinsic Rashba one $\lambda''_{SOC}$ due to an equivalent out-of-plane electric field in each layer, which has finite value at K and will lead to band inversion[5]. For each K valley both of them contributes the mixture of states in sub-orbital space suggested by $\mu_x$. However, only extrinsic Rashba SOC results in the coupling between both sub-orbital and sublattice space. This kind of layer-resolved Rashba SOC can be recognized as a consequence of two layer coupled by antiparallel electric field along the out-of-plane direction, which is not easy to achieve in practice as we require the two layers to be van der Waals separated. It can be recognized that the intrinsic one comes from antiparallel electric field about two sublattice for each layer. For AA stacked bilayer we expect to introduce an effective SOC to induce K band inversion merely from intrinsic Rashba SOC combined interlayer hopping term between sublattice $A_t$ ( $B_t$ ) and $B_b$ ( $A_b$ ). Indeed, there just breaks the particle-hole symmetry of the band structure even though we introduce planar interlayer hopping terms. It is suggested that based on AA stacked bilayer system, we may not achieve that both the interlayer and inter-sublattice hopping term can be independent of $k$ in the Hamiltonian $H(k)$ unless twisting the bilayer system to introduce AA, AB and BA stacking domain regions. It motivates us to be focused on such twisted bilayer system and explore that if it can induce band inversion towards its low energy bands (mini bands near to Fermi level in mini BZ) by an effective SOC.

Studies about bilayer silicene[20], germanene and stanene[21] and also the twisted graphene-silicene heterojunctions[22] have been reported. Experimentally, the scanning tunneling microscopy (STM) images of aligned bilayer silicene show a planar structure[20] in that the two layers are touched through strong chemical bonds, where there loses interlayer $\pi$ bonding to create $\sigma$-like chemical bonds and the electron cloud repulsion will stabilize the planar bilayer structure. According to these observations, we expect that for the twisted structure that the atoms in top layers not all aligned with those in bottom layers, the two layers can be separated by a distance much larger that the low-buckling hold in each layer. And the interlayer bonding, at least the low energy physics, mainly refers to $\pi$ electron in such a van der Waals bilayer junction. It is known to us that the low energy bands in twisted bilayer graphene like system is dominated by the long wave electron states for small twist angle and the low dispersion of the flatness band can be considered as the contribution from effective hopping between stacking regions[23-25]. We then perform the TB calculations with intrinsic SOC included to observe its influence on the low energy band structures.

For our simulation by TB approximation we follow the method as widely used in twisted bilayer graphene (tBLG)[26], the intralayer hopping between site $i$ and site $j$ should be replaced by $t_{ij} = V_1 e^{-\frac{|r|-a_0}{\delta_0}}\left(1-\frac{r_z^2}{|r|^2}\right)$ in Eq. (1) and the interlayer one is replaced by $t_{ij} = V_1 e^{-\frac{|r|-a_0}{\delta_0}}\left(1-\frac{r_z^2}{|r|^2}\right) + V_2 e^{-\frac{|r|-a_0}{\delta_0}}\frac{r_z^2}{|r|^2}$ in Eq. (2). Here, $r = r_i - r_j$ represents the distance

between atomic sites. $a_0$ and $d_0$ denotes the intralayer and interlayer nearest bond length respectively. $\delta_0$ is an empirical decay length of the bonding strength. These parameters we use can be fitted from first-principles calculated band structure without SOC included (Fig. S1). Here, we choose bilayer germanene as a platform and obtain that $V_1 = -1.4$ eV, $V_2 = 0.23$ eV and $\delta_0 = 0.25a$ in which $a$ is the lattice constant of the monolayer. For the SOC strength we choose $\lambda_{SOC} = 3$ meV, $\lambda'_{SOC} = 4$ meV and $\lambda''_{SOC} = 0$ in Eq. (1).

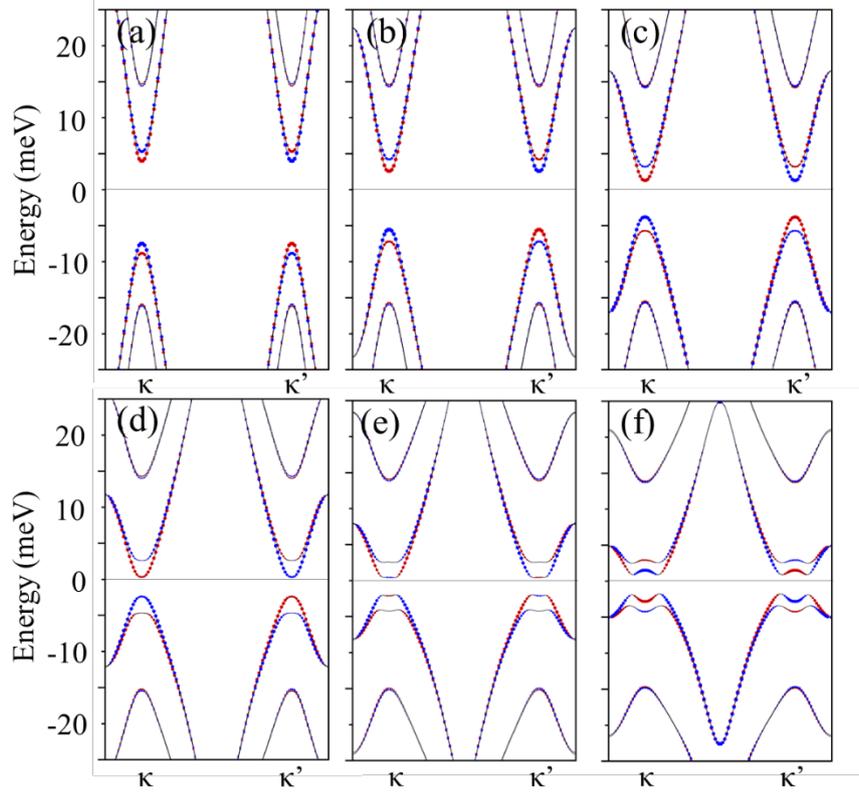

**Figure. 4** Calculated band structures for commensurate superlattice with twist angle $\theta$ about (a) 2.13° (b) 1.89° (c) 1.70° (d) 1.54° (e) 1.41° (f) 1.30°. The Fermi level $E_F$ is set to be at 0.0 eV, denoted as the dashed line above. The red (blue) color shows the calculated positive (negative) spin $\langle s_z \rangle$ polarization on each $k$.

As the twist angle $\theta$ varies, in Fig. 4 we show the calculated band structures near Fermi level $E_F$ and spin $\langle s_z \rangle$ polarization on each $\mathbf{k}$. It is found that there appears band inversion at $\kappa$ and $\kappa'$ valley in mini BZ of our moire superlattice when $\theta$ is about smaller than 1.4° for the case of the TB model parameters at present. Obviously, the smaller twist angle with longer moire period of the superlattice give rise to larger and more distinguished stacking domain regions. And also, there reduces the bandwidth as the effective interlayer coupling within the low energy electron states become stronger, which can be reflected by that the larger value of $m_0$, the larger energy gap at $\kappa$ and $\kappa'$ between the four bands nearest and remote to $E_F$. It is worth noting that this interlayer coupling depends on the twist angle and there exists the band gap at $\kappa$ and $\kappa'$ closing and reopening once, meaning a twist induced topological transition.

It can be inferred that an equivalent layer-resolved Rashba SOC is introduced into the low energy states of twisted bilayer system, which stems from the intrinsic Rashba SOC $\lambda'_{SOC}$ that only couple sub-orbitals and contributes spin-flipping. For the purpose of coupling between sublattices that indeed result in $\kappa$ and $\kappa'$ band inversion as discussed previously, we consider the existence of AB and BA stacking domain regions besides AA regions that effectively couples the sublattice $A_t$ ($B_t$) and $B_b$ ($A_b$) in a way through $H_R$, the intralayer intrinsic Rashba SOC. In a spirit of perturbation, we treat $H_t$, the interlayer hopping within AB and BA regions, as perturbation towards the low energy states and express the effective nonzero matrix element as:

$$\langle A_t + A_b | H_R | B_t - B_b \rangle = \sum_{B'_b} \frac{\langle A_t | H_t | B'_b \rangle}{\Delta} \langle B'_b | - H_R | B_b \rangle + \sum_{A'_b} \langle A_b | H_R | A'_b \rangle \frac{\langle A'_b | H_t | B_t \rangle}{\Delta}$$

(7)

and

$$\langle B_t + B_b | H_R | A_t - A_b \rangle = \sum_{B'_t} \langle B_t | - H_R | B'_t \rangle \frac{\langle B'_t | H_t | A_b \rangle}{\Delta} + \sum_{A'_t} \frac{\langle B_b | H_t | A'_t \rangle}{\Delta} \langle A'_t | H_R | A_t \rangle$$

(8)

Here, Δ represents the energy gap due to the intralayer bonding over the entire regions and interlayer bonding in AA stacking regions, included in $H_0$ and all the eigen states belong to $H_0$. We omit the spin indices and remember that $H_R$ matrix element is among different spin channels. Note that the vertical interlayer hopping in AB and BA regions should give the main contribution, which will be independent of $k$ in Hamiltonian $H(k)$. Thus, the band inversion at κ and κ' can be attributed to an effective Rashba SOC that comes from stacking faults in such moire system. Moreover, from above Eq. (7) and (8) we can see that the sign of the effective electric field in both SOC matrix element $\langle B_b'|-H_R|B_b\rangle$ and $\langle A_b|H_R|A_b'\rangle$ is positive, whereas that in both $\langle B_t|-H_R|B_t'\rangle$ and $\langle A_t'|H_R|A_t\rangle$ is negative. It indicates that, as we expect, there possess equivalent layer-resolved Rashba SOC in the sense that the effective SOC field can be antiparallel for the two layers within AB and BA stacking domain regions.

The nontrivial topological phase in such moire system should possess corresponding topologically protected edge states. For a nanoribbon structure truncated through the AB and BA regions as shown in Fig. 5(b) we calculate the edge band structure. Take 1.35° twisted case as an example, we show its calculated bulk band structure and the corresponding edge band structure in Fig. 5(a), where the red highlighted two bands are the partial filled highest valence band and lowest conduction band. Note that these two edge bands that are found to cross with each other corresponds to the bulk band inversion as a consequence of bulk-edge correspondence. They become topologically protected edge states near Fermi surface which cannot be merged into bulk bands, in agreement with our calculated results of the edge states for bilayer QVHE insulator (Fig 2(b)), where there left one pair of gapless states inside the edge band gap. However, the nonzero valley Chern number here is defined for 'mini' valley index as $C_v = C_\kappa - C_{\kappa'}$. Hence, we would emphasis that it is different from the definition about QVHE case in tBLG system.

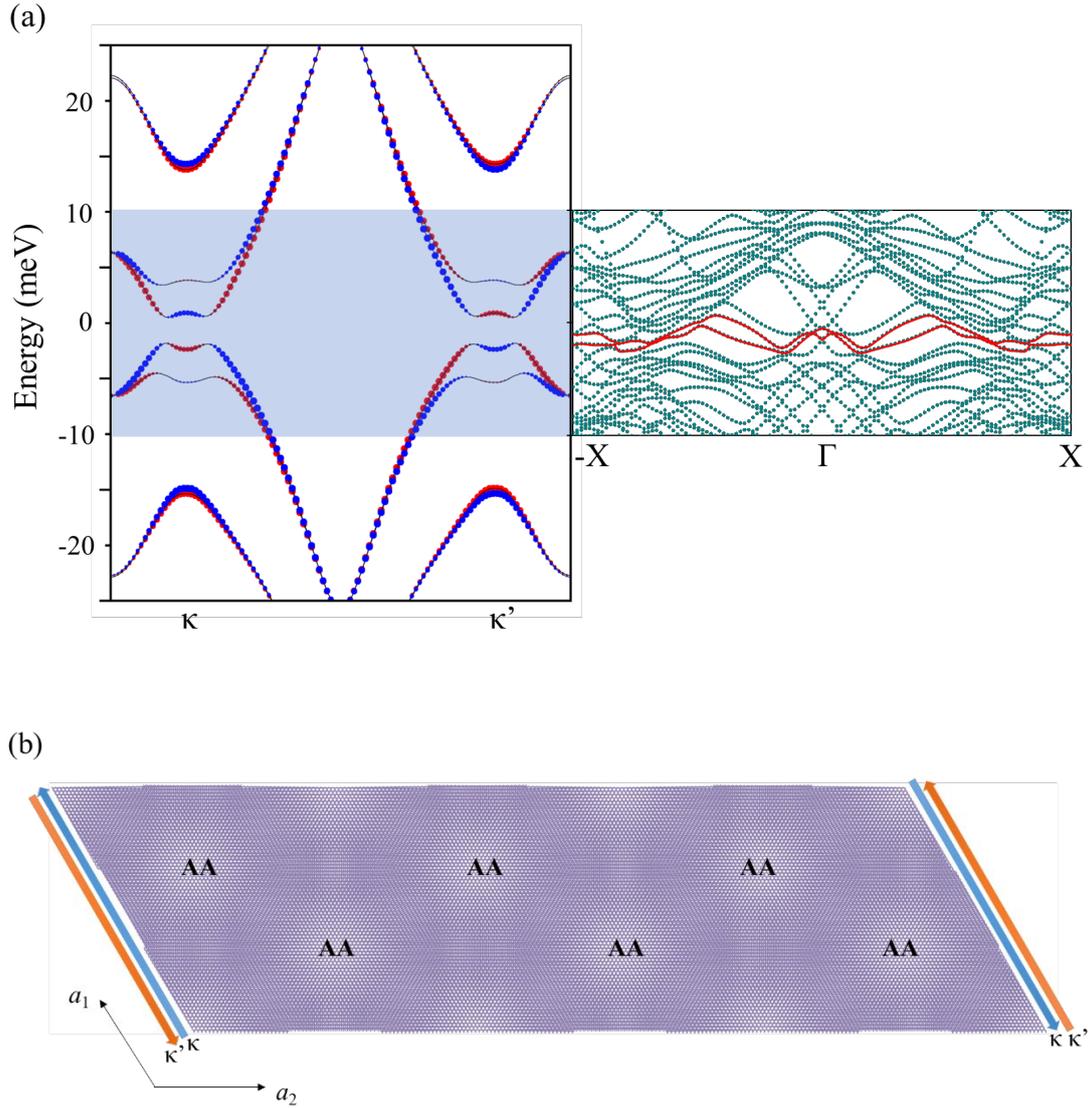

**Figure. 4** (a) Calculated bulk (left panel) and edge (right panel) band structure of the 1.35° twisted case. The energy window for the edge band structure is marked in blue shadow. (b) The nanoribbon structure truncated through the AB and BA regions (64836 atomic sites), where the edge is along $a_2$ direction and the periodic boundary condition is along $a_1$ direction. The valley current at the edge possesses valley-filtering.

In those graphene-based moire system[24,27,28], there emerges quantum anomalous Hall states in each K valley space. Here, the valley index is well defined for graphene atomic cell as long as the intervalley scattering due to the short-range disorder is ignored. Since the two valleys are time-reversal counterparts, the system can exhibit QVHE and valley filtering for its corresponding helical edge modes. However, it may be driven to an intervalley coherent state by strong electron correlation introduced for the flatness bands near Fermi level and then the valley K and K' now is not a good quantum number[28-30]. Different from the flat band case in tBLG, in our case the κ valley in mini BZ is a good quantum number if the intervalley scattering due to a short-range disorder in a moire length scale is absent. Furthermore, note that for our TB calculations, the inter K valley coupling is indeed included so that this valley index is not well defined. Therefore, we need to treat 'mini' valley states as the quantum valley Hall state that appears in the nontrivial topological phase in our definition. And it can be an intra κ valley coherent state even the electron-electron interaction is included.

The 'mini' valley Chern numbers $C_v = C_\kappa - C_{\kappa'}$ of the low energy bands are also associated with valley-contrasting orbital magnetizations, similar to the case of QVHE in tBLG. For each Chern sector in this case[27,28,31], there give rise to a nonvanishing orbital magnetization linearly dependent on the Fermi level variation inside the gap. The QVHE then can be understood by the cyclotron motion formed within the moire superlattice that filters valley states in this way, just similar to the physical picture of the integer quantum Hall effect. In particular there forms current loops that corresponds to each K valley in tBLG system, whose characteristic radius is on the order of a moire cell scale. However, in our case of κ valley states, this order is at least $\sqrt{3}\times\sqrt{3}$ moire cell scale since the corresponding orbital motion forms among moire supercells. And there generates the pure valley current at the samples boundaries, schematically shown in Fig. 4(b), which also possess valley-filtering.

## CONCLUSION

We have shown a quantum valley Hall state in bilayer silicene family in absent of inversion symmetry. It has been found that a layer-resolved Rashba SOC acts as a switch for a nontrivial topological transition, leading the system from two individual QSHE layers to the QVHE phase. Of particular interest is that kind of SOC are introduced by antiparallel electric field for the two layers equivalently and the QVHE can be achieved by such 'antiferroelectric' coupling between layers. Moreover, we find the introduction of different stacking order regions for such bilayer system i.e., twisting the two layers can give rise to the layer-resolved SOC as we expect in an equivalent way. Therefore, a nontrivial mini-bands topology in the moire system has been found accompanied by a topological phase transition induced by interlayer twist. And there generates QVHE for the 'mini' valley states as we define. Such findings of dissipationless valley-filtering in this kind of QVHE insulator extend the potential applications both in valleytronics and twistronics.

# ACKNOWLEDGMENTS

This work was supported by the National Key Research and Development Program of China (Grants No. 2022YFA1402902 and 2021YFA1200700), the National Natural Science Foundation of China (Grants No. 12134003 No. 12188101 No. 11834006), the excellent program in Nanjing University, and Innovation Program for Quantum Science and Technology (No. 2021ZD0301902). National funded postdoctoral researcher program of China (Grant No. GZC20230809), Shanghai Science and Technology Innovation Action Plan (No. 21JC1402000), ECNU Multifunctional Platform for Innovation.

# SUPPORTING INFORMATION

## A. First-principles calculations

For the monolayer germanene, the bond length $a_0$=2.42 Å and buckling angle 106°[5]. The bilayer distance we choose $d_0$=5 Å. The first-principles calculations are performed with density functional theory (DFT) using the projector augmented wave (PAW) method implemented in the Vienna ab initio Simulation Package (VASP)[32]. The exchange-correlation potential is treated in Perdew-Burke-Ernzerhof form[33] of the generalized gradient approximation (GGA-PBE) with a kinetic-energy cutoff of 400 eV. A well-converged 11×11×1 Monkhorst–Pack $k$-point mesh is chosen in self-consistent calculations. The convergence criterion for the electronic energy is $10^{-5}$ eV. In our calculations, we fix all the atomic positions and the dispersion corrected DFT-D2 method[34] is adopted to describe the van der Waals interactions.

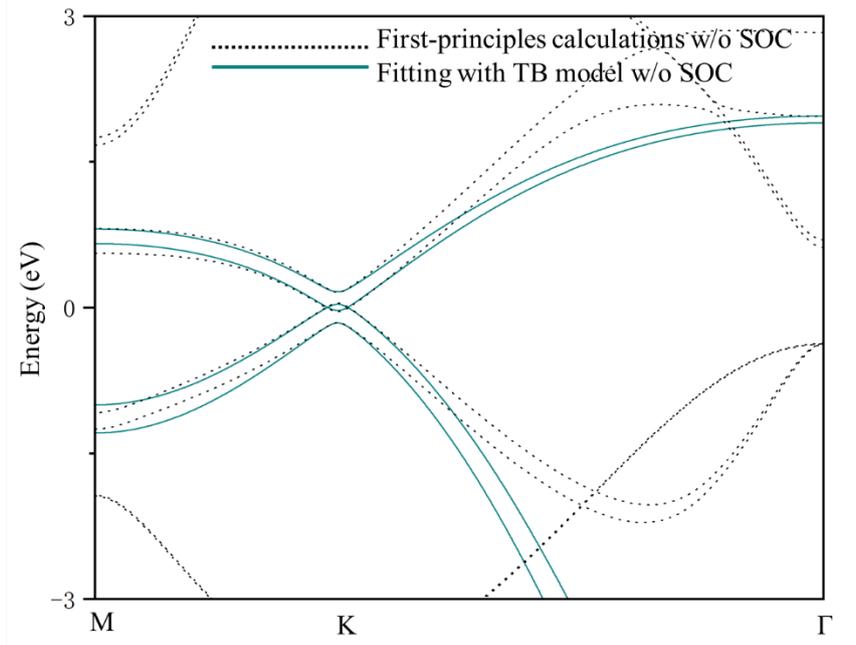

**Figure S1**. For AA stacked bilayer germanene, fitted band structure of TB model (solid line) with first-principles calculated results (dashed line) without SOC included.